\newcommand{\bea}{\begin{eqnarray}}
\newcommand{\eea}{\end{eqnarray}}
\newcommand{\be}{\begin{equation}}
\newcommand{\ee}{\end{equation}}
\newcommand{\ben}{\begin{enumerate}}
\newcommand{\een}{\end{enumerate}}
\newcommand{\bi}{\begin{itemize}}
\newcommand{\ei}{\end{itemize}}
\newcommand{\bmi}[1]{\begin{minipage}{#1 cm}}
\newcommand{\emi}{\end{minipage}}

\def\elabel#1{\label{eq:#1}}
\def\eck#1{\left\lbrack #1 \right\rbrack}

\def\rund#1{\left( #1 \right)}
\def\abs#1{\left\vert #1 \right\vert}

\def\ave#1{\left\langle #1 \right\rangle}

\def\Re{{\cal R}\hbox{e}}
\def\Im{{\cal I}\hbox{m}}

\def\d{{\rm d}}

\def\eps{{\epsilon}}
\def\epss{{\epsilon^{\rm s}}}

\def\esstar{{{\epsilon^{\rm s}}^{*}}}

\def\vt{{\vartheta}}

\def\vt{{\vartheta}}

\def\Real{{\rm I\mathchoice{\kern-0.70mm}{\kern-0.70mm}{\kern-0.65mm}%
  {\kern-0.50mm}R}}
\def\C{\rm C\kern-.42em\vrule width.03em height.58em depth-.02em
       \kern.4em}
\font \bolditalics = cmmib10
\def\bx#1{\leavevmode\thinspace\hbox{\vrule\vtop{\vbox{\hrule\kern1pt
        \hbox{\vphantom{\tt/}\thinspace{\bf#1}\thinspace}}
      \kern1pt\hrule}\vrule}\thinspace}

\def \vc #1{{\textfont1=\bolditalics \hbox{$\bf#1$}}}
{\catcode`\@=11
\gdef\SchlangeUnter#1#2{\lower2pt\vbox{\baselineskip 0pt \lineskip0pt
  \ialign{$\m@th#1\hfil##\hfil$\crcr#2\crcr\sim\crcr}}}
}

\def\ueber#1#2{{\setbox0=\hbox{$#1$}%
  \setbox1=\hbox to\wd0{\hss$\scriptscriptstyle #2$\hss}%
  \offinterlineskip
  \vbox{\box1\kern0.4mm\box0}}{}}

\def\bx#1{\leavevmode\thinspace\hbox{\vrule\vtop{\vbox{\hrule\kern1pt
        \hbox{\vphantom{\tt/}\thinspace{\bf#1}\thinspace}}
      \kern1pt\hrule}\vrule}\thinspace}

{\catcode`\@=11
\gdef\SchlangeUnter#1#2{\lower2pt\vbox{\baselineskip 0pt \lineskip0pt
  \ialign{$\m@th#1\hfil##\hfil$\crcr#2\crcr\sim\crcr}}}
}

\documentclass[referee]{aanew}
\usepackage{graphicx,latexsym}     
\begin{document}
\title{Suppressing the contribution of intrinsic galaxy alignments to 
the shear two-point correlation function}
\author{Lindsay King \& Peter Schneider}
\institute{Institut f{\"u}r Astrophysik und Extraterrestrische
Forschung, Universit{\"a}t Bonn, Auf dem H{\"u}gel 71, D-53121 Bonn,
Germany\\
}
\date{}
\authorrunning{King \& Schneider}
\titlerunning{Suppression of intrinsic alignments in cosmic shear surveys}
\abstract{Cosmological weak lensing gives rise to correlations in the
ellipticities of faint galaxies. This cosmic shear signal depends upon
the matter power spectrum, thus providing a means to constrain
cosmological parameters. It has recently been proposed that {\em
intrinsic} alignments arising at the epoch of galaxy formation can
also contribute significantly to the observed correlations, the
amplitude increasing with decreasing survey depth. Here we consider
the two-point shear correlation function, and demonstrate that
photometric redshift information can be used to suppress the intrinsic
signal; at the same time Poisson noise is increased, due to a decrease
in the effective number of galaxy pairs.  The choice to apply such a
redshift-depending weighting will depend on the characteristics of the
survey in question. In surveys with $\ave z\sim 1$, although the
lensing signal dominates, the measurement error bars may soon become
smaller than the intrinsic alignment signal; hence, in order not to 
be dominated by systematics, redshift information in cosmic shear
statistics will become a necessity. We discuss various aspects of
this.
}
\def\map{{$M_{\rm ap}$}}
\maketitle

\section{Introduction}
The distortion of distant galaxies by the tidal gravitational field of 
intervening matter inhomogeneities has become known as ``cosmic shear''.
Since this lensing signal depends upon the matter power spectrum, it is 
an important cosmological tool, as was proposed in the early 1990s 
by Blandford et al. (1991), Miralda-Escud\'e (1991) and Kaiser (1992). 
This set the scene for further analytic and numerical work 
(eg. Kaiser 1998; Schneider et al. 1998; White \& Hu 2000), taking into 
account the non-linear evolution of the power spectrum which results in 
increased power on small scales (Hamilton et al. 1991; 
Peacock \& Dodds 1996). 

During 2000, four teams announced the first observational detections
of cosmic shear (Bacon et al. 2000; Kaiser et al. 2000; van Waerbeke
et al. 2000; Wittman et al. 2000; Maoli et al. 2001), demonstrating
the feasibility of its study. Various statistics are used to quantify
cosmic shear, and to compare the observations with predictions for
cosmological models. Here we focus on the shear correlation functions;
other measures include the aperture mass statistic (Schneider et
al. 1998) and shear variance (e.g. Kaiser 1992).

Besides its dependence on the large-scale structure in the Universe,
the magnitude of the effect is also sensitive to the redshift
distribution of the galaxies used in the analysis. So far, direct
redshift estimates have not been obtained for the samples
concerned. Typically the mean redshift has been estimated from surveys
of similar depth, and a corresponding redshift probability
distribution has been assumed (e.g. van Waerbeke et al. 2001; Hoekstra
et al. 2002a).  Motivated by the recent interest in obtaining
photometric redshifts for cosmic shear surveys, we consider how to
make use of redshift information, and its impact on the constraints
permitted by the two-point statistic under consideration.
 
In weak lensing analyses, it is assumed that the background galaxies
are randomly oriented so that their mean intrinsic ellipticity
$\ave{\epsilon^{\rm s}}=0$ and any correlation in observed
ellipticities arises from gravitational lensing. The magnitude of any
{\em intrinsic} alignment of the background source population has
recently been the subject of many numerical, analytic and observational
studies (e.g. Croft \& Metzler 2000; Heavens et al. 2000; Crittenden
et al. 2001 (Cr01); Catelan et al. 2001; Mackey et al. 2002; Brown et
al. 2002, Hui \& Zhang 2002). Due to differences in the assumed origin of intrinsic
alignments and the type of galaxies considered, the numerical and
analytic estimates span a couple of orders of magnitude in
amplitude. For example, the analytic model of Cr01 assumes that any
intrinsic signal arises from correlations between the angular momenta
of galaxies, whereas that of Catelan et al. (2001) relies on tidal
shear correlations. Nonetheless, all studies conclude that intrinsic
alignments are more important for shallower surveys (becoming
comparable to the lensing signal for a mean survey redshift ${\ave
z}\sim 0.5$) and that the correlation falls off quite rapidly with
source separation.

The structure of this paper is as follows. In the next section we
outline the relationships between the matter power spectrum and the
lensing correlation functions. The dependence on the redshift
distribution and on cosmology is highlighted.  Correlation functions
are derived from observational data, and in ${\rm Sect.\,3}$ we
describe how practical estimators are used for this purpose.  In ${\rm
Sect.\,4}$ a toy model to account for intrinsic alignments is
developed; the amplitude of the correlation is normalised to that of
Cr01.  The introduction of a weighting factor to minimise any
contribution of intrinsic alignments to the shear correlation function
is considered in ${\rm Sect.\,5}$. In ${\rm Sect.\,6}$ we present the
results of applying such a weighting factor to surveys of different
mean redshift. We finish in ${\rm Sect.\,7}$ with a discussion of the
results and a future perspective of using redshift information in
cosmic shear analysis.

For a review of cosmological weak lensing and the relevant aspects of
 cosmology, see Mellier (1999) and Bartelmann \& Schneider (2001;
 hereafter BS01). In addition, the present status and outlook for
 cosmic shear studies is summarised by van Waerbeke et al. (2002).

\section{Power spectra and lensing correlation functions}
In this section, the relationships between the power spectra and
{\em observable} lensing correlation functions are outlined; throughout, 
we adopt the notation of BS01.

The three-dimensional density fluctuation field $\delta(\vc{r})$ 
at comoving position $\vc{r}$ is a homogeneous and isotropic 
random field, with a two-point correlation function given by
\be
\ave{\delta(\vc{r})\delta(\vc{r'})}\equiv 
C_{\delta\delta}\left(|\vc{r}-\vc{r'}|\right)\;.
\ee
Such a field is described by its power spectrum 
$P_{\delta}(|\vc{k}|)$ ($\vc{k}$ being the comoving wave-vector), which is the 
Fourier transform of the
two-point correlation function. In Fourier space one can write
\be
\ave{\hat\delta(\vc{k})\hat\delta^{*}(\vc{k}')}\equiv
(2\pi)^{3}\delta_{\rm D}(\vc{k}-\vc{k}')P_{\delta}(|\vc{k}|)\;,
\ee
where $x^{*}$ denotes the complex conjugate of $x$, 
$\hat{x}$ denotes its Fourier transform, and 
$\delta_{\rm D}$ is the Dirac delta function.

The effective convergence $\bar\kappa_{\rm eff}(\vc\theta)$ depends upon
the weighted integral of the density contrast along the line of sight 

\begin{equation}
  \bar\kappa_\mathrm{eff}(\vec\theta) =
  \frac{3H_0^2\Omega_{\rm m}}{2c^2}\,
  \int_0^{w_\mathrm{H}}\,\d w\,\bar{W}(w)\,f(w)\,
  \frac{\delta[f(w)\vec\theta,w]}{a(w)}\;,
\label{eq:6.23}
\end{equation}
where $H_0$ and $\Omega_{\rm m}$ are the values of the Hubble
parameter and the density parameter at the present epoch, and $a(w)$ is
the scale factor at comoving distance $w$, normalized such that
$a(0)=1$ today. The horizon distance is denoted by $w_{\mathrm H}$, and $f(w)$ is the
comoving angular diameter distance, so that the comoving separation is 
$f(w)\vc\theta$. The function $f(w)$ depends 
on the spatial curvature, $K$: 
\begin{equation}
  f(w) = \left\{
  \begin{array}{ll}
    K^{-1/2}\sin(K^{1/2}w) & (K>0)\\
    w & (K=0)\\
    (-K)^{-1/2}\sinh[(-K)^{1/2}w] & (K<0) \\
  \end{array}\right.\;;
\end{equation}
later we assume that $K=0$ (i.e. $\Omega_{\rm m}\;+\;\Omega_{\Lambda}=1$). 
The function $\bar{W}(w)$ accounts for 
the sources being distributed in redshift, and consequent differences in 
lensing signal, 
\begin{equation}
  \bar{W}(w) \equiv \int_w^{w_\mathrm{H}}\d w'\,p(w')\,
  \frac{f(w'-w)}{f(w')}~~~\equiv \ave{R(w,w')}\;,
\label{bigw}
\end{equation}
where $p(w')dw'$ is the comoving distance probability distribution for
the sources. The function $R(w,w')=f(w'-w)/f(w')$ is the ratio
of the angular diameter distance of a source at comoving distance $w'$
seen from a distance $w$, to that seen from $w=0$. 
 
The power spectrum $P_{\kappa}(\ell)$ of the effective convergence, or equivalently of the shear  $P_{\gamma}(\ell)$ (see for example BS01), is related to that of the density fluctuations 
through a variant of Limber's equation in Fourier space (Kaiser 1998)
\be
  P_\kappa(\ell) = \frac{9H_0^4\Omega_{\rm m}^2}{4c^4}\,
  \int_0^{w_\mathrm{H}}\,\d w\,\frac{\bar{W}^2(w)}{a^2(w)}\,
  P_\delta\left(\frac{\ell}{f(w)},w\right)\;,
\label{kappow}
\ee
where $\vec\ell$ is the angular wave-vector, Fourier space 
conjugate to $\vec\theta$. 

We see now that $P_{\kappa}$ and $P_{\gamma}$ are intimately related to 
$\Omega_{\rm m}$, and to $P_{\delta}$ which in turn depends on
$\Omega_{\Lambda}$, the shape parameter 
$\Gamma$ and on $\sigma_{8}$, the density fluctuations in spheres of radius 
$8h^{-1}{\rm Mpc}$. In addition, note the dependence upon the redshift 
distribution
of the sources. Access to the cosmological
parameters is provided through the (observable) shear correlation functions,
which we now turn to.  

The shear has two components and can conveniently be represented as
the complex quantity $\gamma\equiv
\gamma_{1}\,+\,{\rm i}\gamma_{2}$. Throughout, we assume the weak lensing
limit so that $|\gamma|\ll 1$. 
Consider a pair of galaxies at position $\vc\vt$ and
$\vc\vt\,+\,\vc\theta$, where the polar angle of their separation
vector $\vc\theta$ is $\phi$. The tangential and cross-components of
the shear for either of these galaxies are
\be
\gamma_{\rm t}=-\Re\eck{\gamma{\rm e}^{-2i\phi}};~~~~\gamma_{\times}=-\Im\eck{\gamma{\rm e}^{-2i\phi}}. 
\ee
Using the notation of Schneider et al. (2002a), the shear correlation functions are defined as
\be
\xi_\pm(\theta)=\ave{\gamma_{\rm t}\gamma_{\rm t}}\pm\ave{\gamma_\times
\gamma_\times}(\theta)=\int_0^\infty
{\d\ell\,\ell\over 2\pi}\,{\rm J}_{0,4}(\ell\theta)\;P_{\kappa}(\ell)\;,
\elabel{corr}\\
\ee
where J$_{n}$ are $n$-th order Bessel functions of the first
kind. $P_\kappa(\ell)$ can be written in terms of the observable 
correlation functions:
\be
P_{\kappa}(\ell)=2\pi\int_0^\infty\d\theta\,\theta\,\eck{\xi_\pm(\theta){\rm J}_{0,4}(\ell\theta)}\;,
\ee
making use of the orthogonality of Bessel functions. From now on we
focus on $\xi_{+}$, since this contains most cosmological information
on the scales of interest.

Another quantity that is often determined in cosmological weak lensing studies is the shear variance inside
a circle of radius $\theta_{\rm c}$, which is also related to the
convergence power spectrum
\be
\ave{\abs{\bar\gamma}^{2}}(\theta_{\rm c})={1\over
2\pi}\int\d\ell\,\ell\,
P_{\kappa}(\ell)\,{4 {\rm J}_{1}^{2}(\ell\theta_{\rm c})\over\left
(\ell\theta_{\rm c}\right)^{2}}\;.
\elabel{Q1}
\ee
This can also be determined directly by placing circular apertures
onto the data field but is affected by any gaps; these may arise, 
for example, due to the need to mask 
regions containing bright stars and their diffraction spikes. However, 
the shear variance can also be obtained from $\xi_{+}$:
\be
\ave{\abs{\bar\gamma}^{2}}(\theta_{\rm c})=\int{\d\vt\,\vt\over\theta_{\rm c}^{2}}\,
\xi_+(\vt)\,S_+\rund{\vt\over\theta_{\rm c}}\;,
\elabel{Q2}
\ee
where van Waerbeke (2000) showed that 
\begin{equation}
  S_+(x) = \left\{
  \begin{array}{ll}
    {1\over\pi}\eck{4\cos^{-1}\rund{x\over 2}-x\sqrt{4-x^2}} & (x\le 2)\\
    0 & (x>2)\\
  \end{array}\right.\;.
\end{equation}

\subsection{Choice of cosmology, power spectrum and source redshift 
distribution}

Unless otherwise stated, our fiducial cosmology is a $\Lambda$CDM
model with $\Omega_{\rm m}=0.3$, $\Omega_{\Lambda}=0.7$ and 
$H_{0}=70\,{\rm km}\,{\rm s}^{-1}\,{\rm Mpc}^{-1}$. 

A scale-invariant ($n=1$, Harrison-Zel'dovich) spectrum of primordial
fluctuations is assumed. Predicting the shear correlation functions
requires a model for the redshift evolution of the 3-D power spectrum.
The fitting formula of Bardeen et al. (1986; BBKS) is used for the
transfer function, and the Peacock and Dodds (1996) prescription for
the evolution in the nonlinear regime. The power spectrum
normalisation is parameterised with $\sigma_{8}=0.9$, and the shape
parameter $\Gamma=0.21$.

As an aside, the differences in predicted shear correlation functions
using the fitting formula of Eisenstein \& Hu (1999) for the transfer
function (rather than that of BBKS) were found to be minimal.

\begin{figure}
\includegraphics[width=15cm]{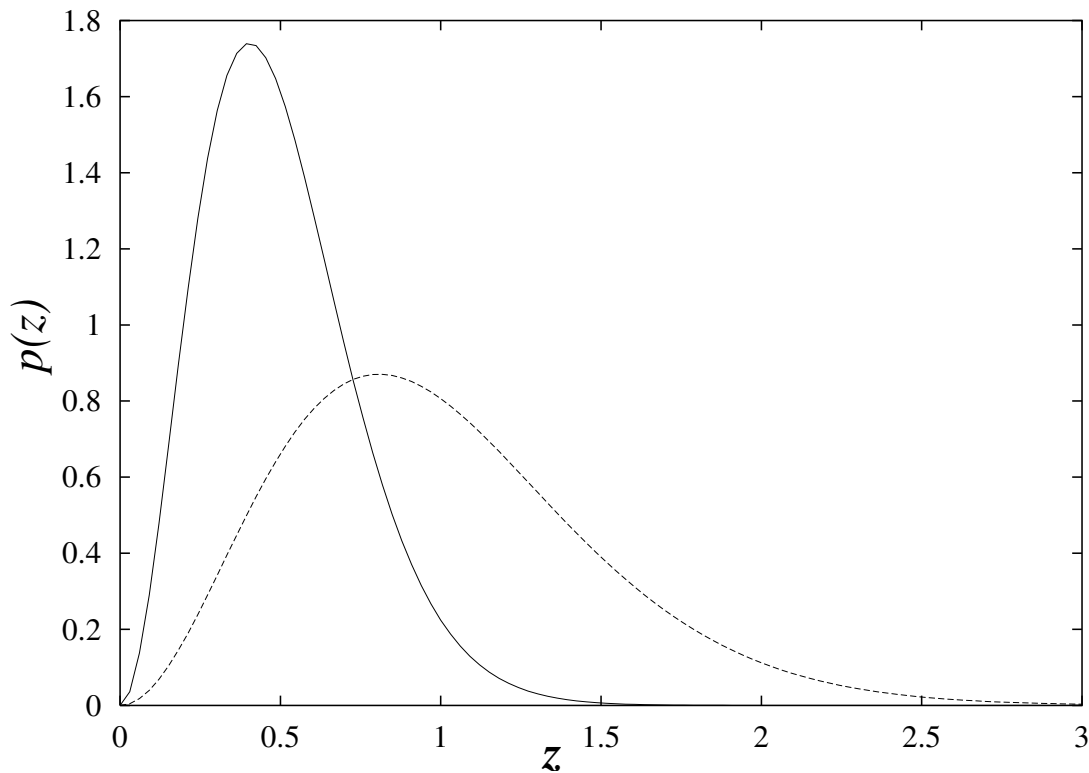}
\caption{
Normalised source redshift distributions, $p(z)$, for 
$\ave z=0.5$ (solid curve) and $\ave z=1.0$ (dashed curve) 
using the parameterisation of Smail et al. (1995).
} 
\label{zb}
\end{figure}

The normalised source redshift distribution is parameterised using the
form suggested by Smail et al. (1995):
\be
p(z){\rm
d}z=\frac{\beta}{z_0\Gamma_{\gamma}\left(3/\beta\right)}\left(\frac{z}{z_0}\right)^{2}{\rm
exp}\left[-\left(\frac{z}{z_0}\right)^{\beta}\right]\,{\rm d}z\;,
\ee
where $\Gamma_{\gamma}(x)$ is the gamma function. We set
$\beta=1.5$ so that $\ave{z}\approx 1.5\,z_{0}$. The value of
$z_{0}$ is adjusted to give different redshift
distributions, which would in practice be obtained from surveys with different
limiting magnitudes. For example, Fig.\,\ref{zb} shows $p(z)$
for $\ave z=0.5$ and $\ave z=1.0$, roughly corresponding to limiting
magnitudes of $I\sim 22$ and $\sim 25$ respectively.
 
\subsection{Some dependencies of $\xi_{+}$}

In this subsection we note some dependencies of $\xi_{+}$ on
cosmological parameters, and on the redshift distribution of the
sources from which it is measured.

\begin{figure}
\includegraphics[width=15cm]{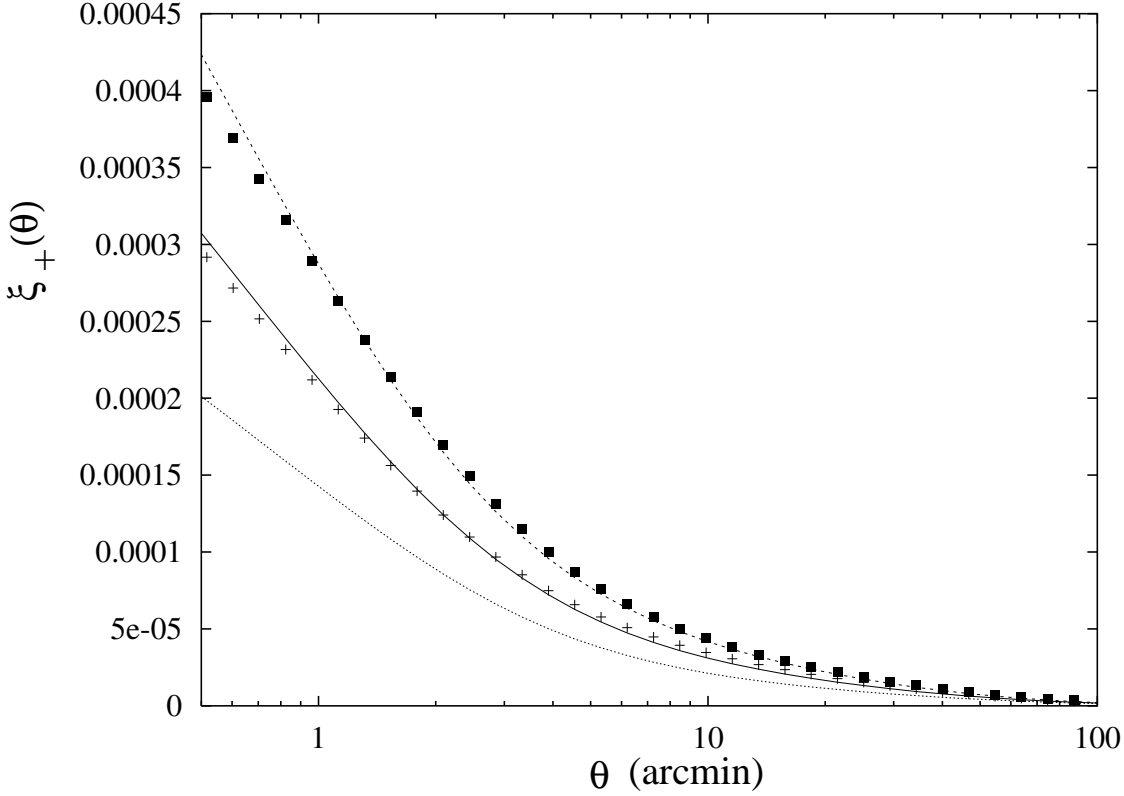}
\caption{Dependence of $\xi_+$ on angular scale $\theta$ for
$\ave{z}=0.8$ (dotted curve), $\ave{z}=1.0$ (solid curve) and
$\ave{z}=1.2$ (short dashed curve). The cosmology for these curves is 
as described in the text, with $\Omega_{\rm m}=0.3$. 
The solid boxes correspond to $\ave{z}=1.0$, now with 
$\Omega_{\rm m}=0.38$. The crosses represent $\ave{z}=1.0$, $\Omega_{\rm m}=0.4$
and $\sigma_{8}=0.78$.} 
\label{eps}
\end{figure}

To illustrate the dependency of $\xi_{+}$ on source 
redshift distribution, Fig.$\,\ref{eps}$ shows 
$\xi_{+}$ for three values of $z_0$, corresponding to 
$\ave{z}=0.8$, 1.0 and 1.2. As expected, the 
amplitude of $\xi_{+}$ is higher for deeper surveys. 
Fig.$\,\ref{eps}$ also highlights a couple of degeneracies:
$\xi_{+}$ for $\ave{z}=1.0$, now 
with $\Omega_{\rm m}=0.38$ is shown. Note the difficulty in
distinguishing this curve from that of the 
fiducial cosmology with a higher mean redshift, $\ave{z}=1.2$.
The well-known degeneracy between $\Omega_{\rm m}$ and $\sigma_{8}$ is
also demonstrated: we show that for 
$\ave{z}=1.0$, taking for instance $\Omega_{\rm m}=0.4$ and 
$\sigma_{8}=0.78$ results in a curve which would 
be difficult to distinguish from that of the fiducial cosmology 
in practice.  

\section{Estimators}

Let us now consider how estimates of the correlation functions can be 
obtained in practice using the distorted images of galaxies.
We begin by considering the case where the only mechanism giving rise
to correlations in galaxy ellipticities is gravitational lensing.
 
In the limit of weak lensing, the complex ellipticity of a galaxy
$\eps_{i}$ at position $\vec\theta_{i}$ is related to its intrinsic
ellipticity $\epss\!\!_{i}$ and to the gravitational shear $\gamma(\vec\theta_{i})$ through
\begin{equation}
\eps_{i}=\epss\!\!_{i}+\gamma(\vec\theta_{i})\;.
\label{uleps}
\end{equation}
We use the definition of ellipticity from Bonnet \& Mellier
(1995) where for elliptical isophotes with axis ratio $b/a\le 1$, 
$|\eps|=(1-b/a)(1+b/a)^{-1}$. Further, under this definition $\ave{\eps}\approx\gamma$ if
the unlensed source population is randomly oriented.
For simplicity we will use the notation
$\gamma_{i}\equiv\gamma(\vec\theta_{i})$. Using (\ref{uleps}) it
follows that the expectation value
\begin{equation}
\ave{\eps_{i}\eps_{j}^{*}}\equiv\ave{\left(\epss\!\!_{i}+\gamma_{i}\right)\left(\esstar\!\!\!\!_{j}+\gamma^{*}_{j}\right)}\equiv\ave{\gamma_{i}\gamma_{j}^{*}}+\ave{\epss\!\!_{i}\esstar\!\!\!\!_{j}}\;,
\end{equation}
and in the absence of intrinsic correlations the final
term vanishes. In this case, any correlation between observed galaxy 
ellipticities results from weak lensing.
 
As discussed in Schneider et al. (2002b),
the correlation function can be estimated in bins of
width $\Delta\vt$, centred on $\vt$, and a weight function can be
defined such that $\Delta_{\vt}(\phi)=1$ for 
$\vt-\Delta\vt/2<\phi\leq\vt+\Delta\vt/2$  and is zero otherwise. An estimator for the correlation function $\xi_{+}(\vt)$ is given by 
\begin{equation}
\hat\xi_{+}(\vt)=\frac{\sum_{ij}{\mathcal W}_{i}{\mathcal W}_{j}
\Delta_{\vt}\left(|\vec\theta_{i}-\vec\theta_{j}|\right)
\left(\eps_{i}\eps_{j}^{*}\right)}{N_{\rm p}(\vt)}
\;,~~~N_{\rm p}(\vt)=\sum_{ij}{\mathcal
W}_{i}{\mathcal
W}_{j}\Delta_{\vt}\left(|\vec\theta_{i}-\vec\theta_{j}|\right)\;,
\label{est1}
\end{equation}
where ${\mathcal W}_{i}$ and ${\mathcal W}_{j}$ are weights (depending on the
reliability of the ellipticity measurement, for example) 
and $N_{\rm p}(\vt)$ is the effective 
number of pairs in that bin. In the absence of intrinsic correlations, 
since $\ave{\eps_{i}\eps^{*}_{j}}\equiv
\xi_{+}(|\vc\theta_{i}-\vc\theta_{j}|)$, it follows that 
$\hat\xi_{+}(\vt)$ is an unbiased estimator of the shear correlation 
function of the lensing signal $\xi_{+}(\vt)$. 

However, the existence of intrinsic alignments would 
imply that $\ave{\epss\!\!_{i}\esstar\!\!\!\!_{j}}\neq 0$. The
estimator $\hat\xi_{+}(\vt)$ is no longer an unbiased estimator of
$\xi_{+}(\vt)$: rather it includes a contribution from correlated
intrinsic (source) ellipticities, for which the two-point correlation 
function will be denoted by $\xi_{+}^{\rm int}(\vt)$. 
In order to minimise the impact of intrinsic correlations on the
interpretation of the weak lensing signal, we need an estimate of
the amplitude of  $\xi_{+}^{\rm int}(\vt)$, and a scheme to give less
weight to galaxy pairs where the intrinsic correlation is likely to be 
high.  

\section{Intrinsic alignments}
Correlations between the intrinsic ellipticities of neighbouring
galaxies are expected to arise during their formation, due to
similarities in the tidal gravitational field in which they form. 
As noted in the introduction, a number of numerical and analytical
studies have sought to quantify the amplitude of this correlation 
as a function of angular separation and redshift.
For our purposes, we need a toy model to roughly predict the amplitude of 
intrinsic correlations expected. Since there is 
disagreement on the dominant mechanism responsible for intrinsic 
alignments, and on its subsequent enhancement or ``washing out'',
use of a detailed model is not warranted here.

Let us consider a close pair of galaxies with comoving distances
$w_{i}\approx w_{j} \approx w$  and angular separation $\theta$. 
Their comoving separation $r$ is given by
\begin{equation}
r^{2}=(\Delta w)^{2}+ f^{2}(w)(\theta)^{2}\;,
\end{equation}
where $\Delta w \approx w_{i}-w_{j}$. 

Denoting the (3-D) two-point density correlation function by 
$\xi_{\delta}(r)$, one might
expect that the two-point correlations in intrinsic ellipticity 
$\propto \xi_{\delta}^{2}(r)$. Indeed, this
dependence corresponds to the asymptotic behaviour at large
separations derived by Cr01, whose normalisation we later
adopt. The two-point density correlation function is the Fourier
transform of the power spectrum, so the power spectrum used to predict
the lensing correlation functions described in Sect.\,2 could be
used. However, since Cr01 focus on a
$\xi_{\delta}(r)\propto 1/r$ behaviour, and because of its analytic simplicity, we also use this form. 

Next, this 3-D intrinsic correlation function has to be projected into
an angular intrinsic 
ellipticity correlation function, integrating over the distance
distribution of galaxies and taking into account the (observed) galaxy 
two-point correlation function $\xi_{\rm gg}(r)$. 
Note that we do not consider the 
individual tangential and cross components of the correlation functions of the intrinsic
ellipticities. Rather, we consider the sum of these quantities, which
corresponds to $\xi_{+}$ in the notation of Schneider et al. (2002a; 
note that the notation of Cr01 differs here). 
We obtain
\begin{equation}
\xi_{+}^{\rm int}(\theta)\propto \frac{\int{\rm d}w_{1}\,p(w_{1})
\int{
{\rm d}w_{2}\,p(w_{2})[1+\xi_{\rm gg}(r)]
\left[\xi_{\delta}(r)\right]^{2}}}
{\int{\rm d}w_{1}\,p(w_{1})
\int{{{\rm d}w_{2}\,p(w_{2})[1+\xi_{\rm gg}(r)]}}}\;,
\end{equation}
where $p(w){\rm d}w$ is the galaxy distance distribution, which we
take to be of the form given in Sect.\,2. 

Given that the scale over which intrinsic alignments operate is much 
smaller than the distances to the galaxies, the integral above can be
recast as

\begin{equation}
\xi_{+}^{\rm int}(\theta)\propto \frac{\int{
{\rm d}w \;p^{2}(w)\;\int{{\rm d}(\Delta w)\; [1+\xi_{\rm gg}(r)]\;\left[\xi_{\delta}(r)\right]^{2}}}}
{1\;+\;\int{{\rm d}w\; p^{2}(w)\;\int{{\rm d}(\Delta w)\;\left[\xi_{\rm gg}(r)\right]}}}\;.
\end{equation}

The two-point galaxy correlation function is taken to be 
$\xi_{\rm gg}(r)=(r/4.3\;h^{-1}{\rm Mpc})^{-1.8}$, constant in
comoving separation. Evolution is
ignored for both the galaxy and density correlation functions, in view
of the weak evolution that is seen in galaxy samples out to $z\sim 3$
(e.g. Porciani \& Giavalisco 2002). 
The normalisation of $\xi_{+}^{\rm int}(\theta)$ is set at an angular scale of
$10\arcmin$ and for $\ave{z}=1$, using the value obtained by
Cr01; Mackey et al. (2002) point out that on very small spatial scales
($\sim 1\,h^{-1}\,{\rm Mpc}$) non-linear clustering effects may erase
any initial alignments, so we perform the normalisation well above
this scale.  We end up with a simple model 
for $\xi_{+}^{\rm int}(\theta)$, which depends upon the redshift 
distribution of sources. 

\section{Inclusion of a redshift dependent weighting factor}

An important difference between ellipticity correlations arising from lensing 
and those due to intrinsic alignment is that the latter only has a 
significant amplitude over a relatively small range in redshift.
For a given galaxy pair, the lensing correlations depend on the integrated 
effect of the matter along the line of sight out to the redshift of the 
closer galaxy. 

In order to minimise the intrinsic signal, giving less weight to galaxy 
pairs where the redshift difference is small seems like a plausible step.
How effective would such a weighting be, how would the lensing correlation be 
changed and how would the noise on our estimate of the correlation function 
increase? 

Comparison of the observed correlation functions with the foregoing 
expressions requires one to adopt a source redshift distribution; 
in practice, what will become routinely available are
photometric redshift estimates. The true redshift of galaxy $i$ will
be denoted by $z_{i}$ and the photometric redshift estimate by 
$\bar z_{i}$, with equivalent notation for distances (i.e. $w_{i}$ and $\bar w_{i}$). The probability density to have sources at comoving distance $w_{i}$ and to have photometric distance estimates $\bar w_{i}$ will be denoted by 
$p(w_{i})$ and $p(\bar w_{i})$ respectively, and $p(w_{i},w_{j})$ is 
the joint probability to have sources at $w_{i}$ {\em and} at $w_{j}$ . For conditional probabilities, $p(w_{i}|{\bar w_{i}})$ represents the probability that the true source distance is $w_{i}$ given that the photometric distance estimate is $\bar w_{i}$. Similarly, $p({\bar w_{i}}|w_{i})$ is the probability that a photometric distance estimate $\bar w_{i}$ will be obtained, given that the true distance is $w_{i}$. 
 
\subsection{Estimator}
A practical estimator for the lensing correlation function should
minimise the relative signal from intrinsic alignments:
\begin{equation}
\hat\xi_{+}(\vt)=\frac{\sum_{ij}{\mathcal W}_{i}{\mathcal
W}_{j}{\cal
Z}_{i,j}\Delta_{\vt}\left(|\vec\theta_{i}-\vec\theta_{j}|\right)\left(\eps_{i}\eps_{j}^{*}\right)}{N_{\rm
p}(\vt)}\;,~~~N_{\rm p}(\vt)=\sum_{ij}{\mathcal
W}_{i}{\mathcal W}_{j}{\cal
Z}_{i,j}\Delta_{\vt}\left(|\vec\theta_{i}-\vec\theta_{j}|\right)\;;
\label{est}
\end{equation}
${\cal Z}_{i,j}\equiv{\cal Z}(\bar{z_{i}},\bar{z_{j}})$ is a weighting
factor which could be of the form
\begin{equation}
{\cal Z}(\bar{z_{i}}, \bar{z_{j}})=1-{\rm exp}\left[-\frac{(\Delta \bar{z})^2}{2\;\sigma_{{\cal Z}}^2}\right]\;,
\label{weight}
\end{equation}
where $\Delta \bar{z}$ is the difference between the photometric
redshift estimates of the
galaxies in a pair, and $\sigma_{{\cal Z}}$ is a factor that controls the extent of the region in 
$\Delta \bar{z}$ space over which galaxy pairs are down-weighted. This would 
be chosen to reflect photometric uncertainty, which is typically much larger 
than the range over which intrinsic correlations operate.
We will neglect the weights $\cal W$ in what follows (since these
would be determined from the data in practice), and see how this
estimator performs in Sect.\,6.

\subsection{Lensing correlation function}
We first consider what effect a redshift dependent weighting factor
 has on the lensing induced component of the correlation function,
 $\xi_{+}(\theta)$. The addition of such a weighting factor modifies
 the effective source distance 
distribution, and a distance correlation is introduced.  To make this
clear, consider the shear correlation function for particular source
distances $w_{i}$ and $w_{j}$ i.e. $\xi_{+}(\theta; w_{i},w_{j})$, 
which would correspond to sources located on two sheets 
(Schneider et al 2002a); 
\bea
&&\xi_{+}(\theta;w_i,w_j)
=
{9 H_0^4 \Omega_0^2\over 4 c^4}
\int_0^{w_{i,j}} {\d w\over a^2(w)}\nonumber \\
&&\times
R(w,w_i)\,R(w,w_j) 
 \int{\d \ell\,\ell\over (2\pi)}\,P_\delta\rund{{ \ell\over f(w)},w}\,{\rm
J}_0(\ell\theta) \,,
\label{eqnshe}
\eea
where the upper limit $w_{i,j}$ in the outer integral is the minimum 
of $w_{i}$ and $w_{j}$.
What is important to note is the dependence on the 
product $R(w,w_i)\,R(w,w_j)$. When (\ref{eqnshe}) is averaged over the source
distance distribution, Schneider et al. (2002a) obtain
\bea
\xi_{+}(\theta)=
{9 H_0^4 \Omega_0^2\over 4 c^4}
\int_0^{w_{\rm H}} {\d w\over a^2(w)}\;
\int{\d \ell\,\ell \over (2\pi)}\,P_\delta\rund{{ \ell\over f(w)},w}
{\rm
J}_{0}(\ell\theta)
\ave{R(w,w_i)\,R(w,w_j)} \;,
\label{G56}
\eea
where the angular brackets in $\ave{R(w,w_{i})\,R(w,w_{j})}$ denote
averaging over the probability distribution $p(w_{i},w_{j})$.  If
$p(w_{i},w_{j})$ is the probability density to have sources at
comoving distances $w_{i}$ and $w_{j}$, in the absence of intrinsic
source correlations one has that $p(w_{i},w_{j})\equiv
p(w_{i})p(w_{j})$. Then, $\ave{(R(w,w_{i})R(w,w_{j})}\equiv\bar
W^{2}(w)$, as in in (\ref{kappow}). The addition of the weighting
factor ${\cal Z}({\bar w}_{i}, {\bar w}_{j})$ changes the probability
density of the redshifts of pairs included in the estimator (\ref{est})
thus
\begin{equation}
p(w_{i},w_{j})=\frac{\int_{0}^{w_{H}}{\rm d}{\bar
w}_{i}\;{\cal X}\int_{0}^{w_{H}}{\rm d}{\bar w}_{j}\;{\cal Y}\;{\cal
Z}(\bar{w}_{i},\bar{w}_{j})}
{
\int_{0}^{w_{H}}{\rm
d}{\bar w}_{i}\;\int_{0}^{w_{H}}{\rm d}{\bar
w}_{j}\;{\cal
Z}(\bar{w}_{i},\bar{w}_{j})
\int_{0}^{w_{H}}{\rm
d}w_{i}\;{\cal X}\int_{0}^{w_{H}}{\rm
d}w_{j}\;{\cal Y}}\;,
\end{equation}
where ${\cal X}\equiv p({\bar w}_{i})\,p(w_{i}|{\bar w}_{i})$ and 
${\cal Y}\equiv p({\bar w}_{j})\,p(w_{j}|{\bar w}_{j})$, 
the inclusion of the denominator ensuring that the probability
distribution is normalised. Using $p({\bar w_{i}}|w_{i})\;p(w_{i})=p(w_{i}|{\bar w_{i}})\;p({\bar w_{i}})$ (and similarly for galaxy $j$) we have that the expectation value of 

\begin{equation}
\ave{R(w,w_{i})R(w,w_{j})}=
\frac
{
\int_{w}^{w_{H}}{\rm d}w_{i}\,R(w,w_{i})\,
\int_{w}^{w_{H}}{\rm d}w_{j}\,R(w,w_{j})\,
\int_{0}^{w_{H}}{\rm d}{\bar w}_{i}\;{\cal X}'\,
\int_{0}^{w_{H}}{\rm d}{\bar w}_{j}\;{\cal Y}'
}
{
\int_{0}^{w_{H}}{\rm d}w_{i}\,
\int_{0}^{w_{H}}{\rm d}w_{j}\,
\int_{0}^{w_{H}}{\rm d}{\bar w}_{i}\;{\cal X}'\,
\int_{0}^{w_{H}}{\rm d}{\bar w}_{j}\;{\cal Y}'
}\;,
\label{rr}
\end{equation}
where ${\cal X}'\equiv p(w_{i})\;p({\bar w}_{i}|w_{i})$ ~and~ ${\cal
Y}'\equiv p(w_{j})\;p({\bar w}_{j}|w_{j}){\cal Z}({\bar w}_{i}, {\bar w}_{j})$.
Substitution of (\ref{rr}) into (\ref{G56}) gives the lensing 
two-point correlation $\xi_{+}(\theta)$. This depends upon the
accuracy of photometric redshift estimates, and upon ``filtering"
using the weighting factor $\cal Z$.

\subsection{Intrinsic correlation function}

Now consider how accounting for photometric redshifts and applying the same 
weighting factor ${\cal Z}({\bar w}_{i}, {\bar w}_{j})$ to the prescription for the intrinsic correlation function adjusts the intrinsic correlation function:

\begin{equation}
\xi_{+}^{\rm int}(\theta)= \frac
{
\int_{0}^{w_{H}}{\rm d}w_{i}\,
\int_{0}^{w_{H}}{\rm d}w_{j}\,
\int_{0}^{w_{H}}{\rm d}{\bar w}_{i}\;{\cal X}'\,
\int_{0}^{w_{H}}{\rm d}{\bar w}_{j}\;{\cal Y}'\,
\left[1+\xi_{\rm gg}(r)\right]\;\left[\xi_{\delta}(r)\right]^{2}}
{
\int_{0}^{w_{H}}{\rm d}w_{i}\,
\int_{0}^{w_{H}}{\rm d}w_{j}\,
\int_{0}^{w_{H}}{\rm d}{\bar w}_{i}\;{\cal X}'\,
\int_{0}^{w_{H}}{\rm d}{\bar w}_{j}\;{\cal Y}'\,
[1+\xi_{\rm gg}]\;}\;,
\end{equation}
where ${\cal X}'$ and ${\cal Y}'$ are defined as below (\ref{rr}).
The $\xi_{\rm gg}(r)$ term in the denominator can safely be ignored,
leaving us with the same denominator as (\ref{rr}). We note that the
galaxy correlation function $\xi_{\rm gg}$ should in principle also
enter the expressions for the shear; however, as shown in Schneider et
al.\ (2002a), this modifies the resulting cosmic shear signal only by
a very small fraction and has therefore been ignored throughout.

\section{Results}
\begin{figure}
\includegraphics[width=15cm]{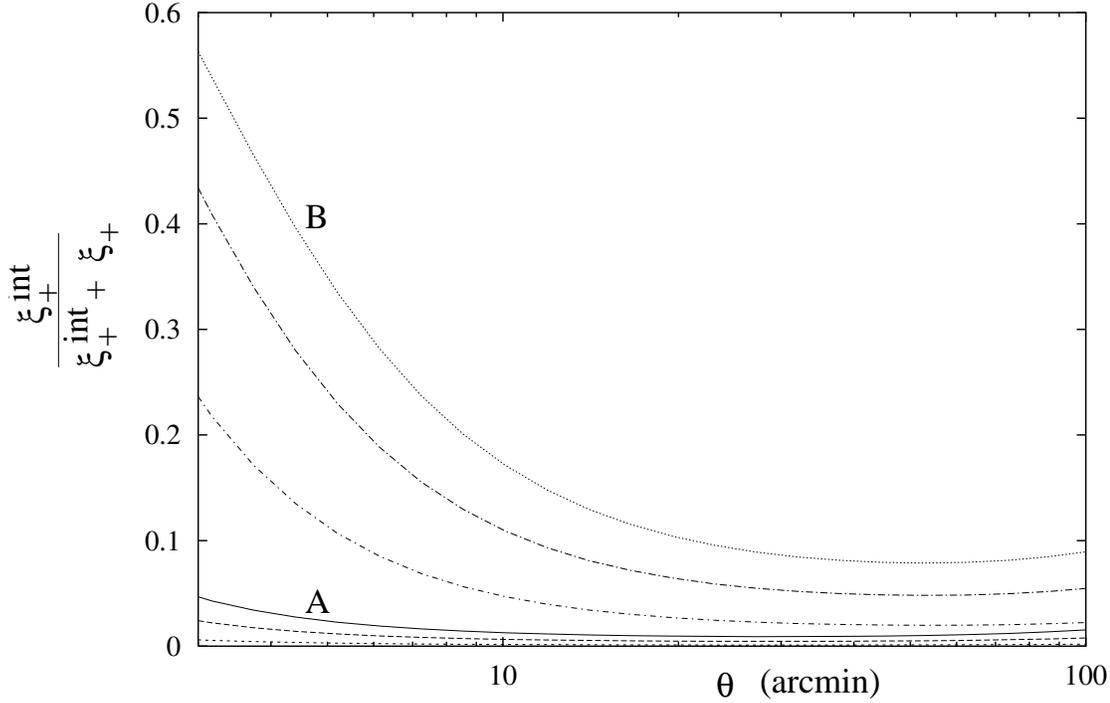}
\caption{
The ratio of the intrinsic correlation function $\xi_{+}^{\rm int}$ to the
combined lensing and intrinsic correlation function
 $(\xi_{+}+\xi_{+}^{\rm int})$ as a function of angular scale $\theta$.
For all cases, the photometric redshift error is assumed to be Gaussian distributed,
with dispersion $\sigma_{\rm phot}=0.1$.
The lower set of three curves are for $\ave{z}=1.0$, with the curve
marked ``A" corresponding to the case where there is no redshift
dependent weighting factor, $\cal Z$. The curve below this is for 
$\sigma_{\cal Z}=0.1$ and the lowermost curve is for $\sigma_{\cal Z}=0.4$.  
The upper set of three curves are for a lower mean redshift
$\ave{z}=0.5$, with the curve marked ``B" being the case where there is
no weighting factor. As for the higher mean redshift case, 
the other two curves correspond to $\sigma_{\cal Z}=0.1$ and $\sigma_{\cal Z}=0.4$ respectively. 
} 
\label{weis1}
\end{figure}

The error in each of our photometric redshift estimates is assumed to
be a Gaussian of dispersion $\sigma_{\rm phot}=0.1$ in $z$, typical of
the accuracy obtained using codes such as hyperz, which adopts a
standard spectral energy distribution fitting procedure (Bolzonella et
al. 2000), provided that photometric information is available in a
sufficient number of wavelength bands.  The ratio of the intrinsic
signal to the total (lensing plus intrinsic) signal for the weighting
factor given by (\ref{weight}) and for various widths of the weighting
function, $\sigma_{\cal Z}$, is shown in Fig.\,\ref{weis1}.  Two
redshift distributions are shown, one with $\ave{z}=1.0$ and the other
with $\ave{z}=0.5$. Note that the contamination from intrinsic
alignments is much more dramatic in the lower redshift case, but that
this can be largely suppressed by down-weighting pairs of galaxies
with similar photometric redshifts.  The higher mean redshift case is
perhaps more typical of surveys suitable for cosmic shear
studies. Although the fractional contribution from the intrinsic
alignment signal is low, it can be reduced.

\begin{figure}
\includegraphics[width=15cm]{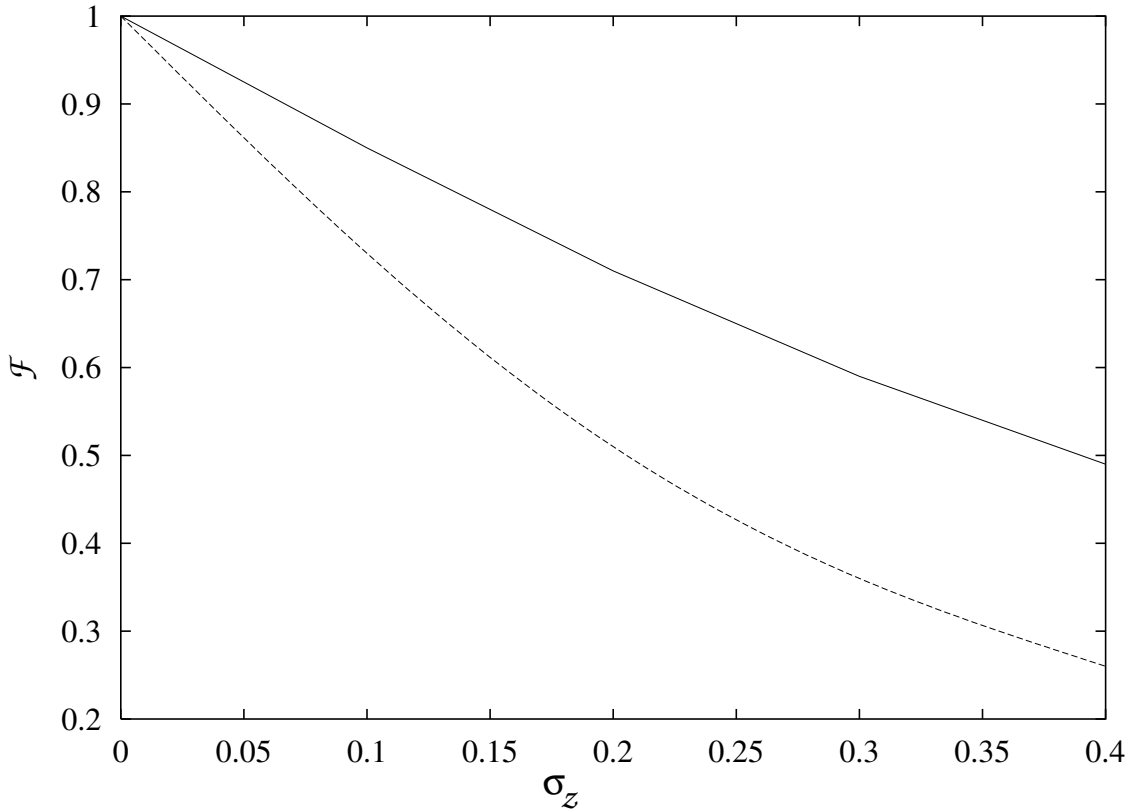}
\caption{
The effective fraction of pairs remaining ($\cal F$) as a
function of $\sigma_{\cal Z}$ (filter width) for $\ave{z}=1.0$ (solid curve) and $\ave{z}=0.5$ (dashed curve). 
} 
\label{norm}
\end{figure}

\begin{figure}
\includegraphics[width=15cm]{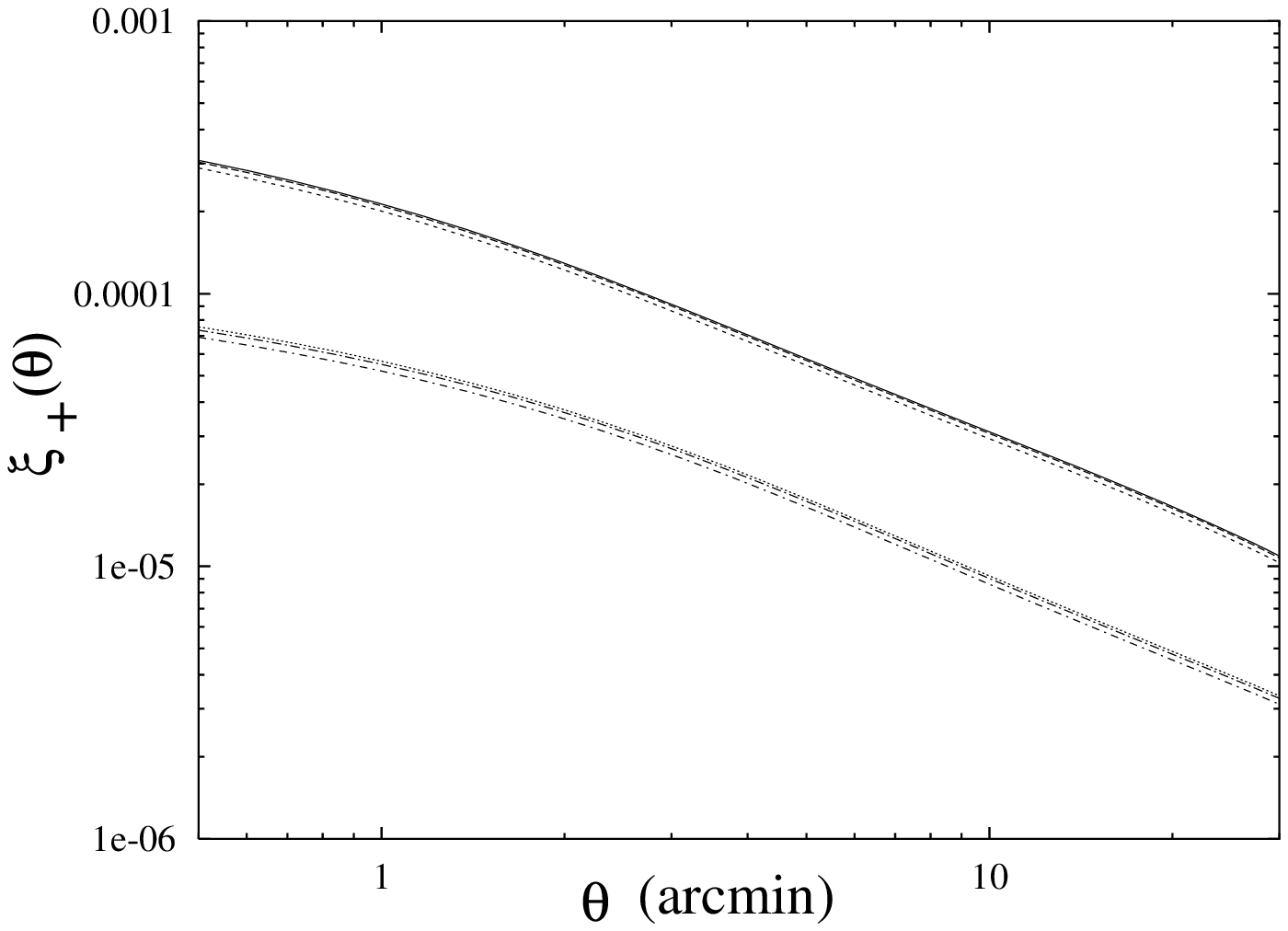}
\caption{
The lensing correlation function $\xi_{+}(\theta)$  as a
function of $\theta$, for ${\ave z}=1$ (upper curves) and for  
${\ave z}=0.5$ (lower curves). Within each set of three curves, the
uppermost curve is for no weighting, the middle curve is for
$\sigma_{\cal Z}=0.1$ and the lowermost curve is for $\sigma_{\cal Z}=0.4$.
} 
\label{diff}
\end{figure}

By down-weighting pairs depending on their photometric redshift
estimates, the number of pairs is also effectively reduced and hence
the noise is increased. For the higher (lower) redshift case, the
solid (dashed) curve of Fig.\,\ref{norm} shows the fraction of pairs
remaining (${\cal F}$) as a function of the filter width $\sigma_{\cal Z}$. 
Although the noise is increased, the expectation value
$\ave{R(w,w_{i})\,R(w,w_{j})}$ is rather insensitive to $\sigma_{\cal
Z}$ and hence the change in $\xi_{+}(\theta)$ is also
small (see Fig.\,\ref{diff}). 

\section{Discussion and Conclusions}

Recent studies of intrinsic ellipticity correlations have shown that
the intrinsic signal can be comparable to (or exceed) the lensing
signal in the case of shallow cosmic shear surveys (Jing 2002). We
investigated to what extent the contribution of galaxy pairs which are
likely to have larger intrinsic correlations can be suppressed, using
photometric redshift information and a simple redshift dependent
weighting factor.  Such a process also results in a reduction in the
effective number of pairs and hence an increase in the noise, which
scales roughly as $1/{\sqrt{N_{\rm p}}}$. The width of the weighting
factor, parameterised by $\sigma_{\cal Z}$, controls the fraction of
pairs remaining. The photometric accuracy determines the precision
with which galaxies at similar redshifts can be identified, hence the
residual contribution from intrinsic alignments.

The reduction of the effective number of pairs is not the only
relevant consideration for the noise in the determination of the
two-point correlation function. Depending on the survey geometry,
cosmic variance can become the dominant contribution to the covariance
of the correlation function for larger angular separations. In
particular, for a compact survey geometry, cosmic variance of the
shear field dominates over shot noise for angular scales larger than a
few arcminutes (e.g., Schneider et al.\ 2002b). Hence, for those
scales the reduction of the effective number of pairs becomes of
little relevance.

Intrinsic alignments have also been invoked as a possible source of
the so-called B-mode contributions seen in cosmic shear surveys (van
Waerbeke et al.\ 2001, 2002; Hoekstra et al.\ 2002b). Such B-modes are
not expected from lensing effects and thus are usually interpreted as
a remaining systematics; whereas galaxy correlations can in principle
generate a B-mode contribution (Schneider et al.\ 2002a), its
amplitude is very small. If the B-mode is due to an intrinsic
alignment of galaxies, then the use of the redshift filter as
discussed here should strongly suppress its relative contribution.

Choosing to apply such a weighting factor will be governed by the
details of the survey. Such considerations include that the impact of
intrinsic alignments is more dramatic for surveys of lower mean
redshift or depth.  Another important factor is the accuracy of
photometric redshift estimates, which depends upon the combination and
characteristics of filters used for the survey - see e.g. Wolf et
al. (2001) for a discussion of photometric redshift estimate
performance in the context of filter sets. Further, the size of the
error bars associated with the practical determination of the
two-point correlation function will also play a part in the decision.
For deep surveys, with our assumed photometric errors, in order to
decrease the fractional contribution of $\xi_{+}^{\rm int}$ by a few
percent, a reduction of several tens of percent in the number of
effective pairs may result. However, removal of the intrinsic
alignment systematic becomes increasingly important as the
experimental error bars in surveys become smaller.  In shallower
surveys, the contamination from intrinsic alignments is much more
pronounced, and in this case down-weighting close galaxy pairs is more
effective. Given real data, one might consider a more elaborate
weighting scheme that is a function of not only the difference in
photometric redshifts, but also giving more weight to higher redshift
pairs. Furthermore, the weight factor ${\cal Z}_{i,j}$ need not be
chosen as a function of the estimated photometric redshift only, but
can be constructed such that it depends on the full redshift
probability distribution as estimated by photometric redshift
techniques. For example, a natural choice would be to have ${\cal
Z}_{i,j}$ depending on the probability that the two galaxies $i,j$ lie
within a narrow redshift interval $\Delta z$, as estimated from their
individual $z$-distributions. The calculation of the expectation value
of the corresponding shear estimator is then slightly more
complicated, but can be done for each survey at hand.

Throughout we have assumed that photometric redshift estimates will be
available for each of the source galaxies, and that the dispersion in
these estimates is independent of magnitude or redshift. In practice,
photometric redshift estimates are reliable down to a certain limiting 
magnitude, depending on the survey characteristics and the relative
depths of the filters used. Particularly in a deep survey, this
magnitude limit may exclude many of
the fainter and smaller galaxies that are likely to be in the
high-redshift tail of the redshift distribution, and to make a
substantial contribution to the lensing signal. Requiring photometric
redshift estimates could in fact lead to an increase in the Poisson
noise of such a survey by reducing the usable number of pairs beyond
the effect shown in Fig.\,4, and in addition the reduction in mean redshift
could increase the relative contribution from intrinsic alignments. 
One possibility to counteract such affects might be to include
fainter galaxies where no photometric redshift estimates are
available, assuming that these are at higher redshifts where intrinsic
alignment is less of an issue.

The use of cosmic shear surveys as a tool to place constraints on the
3-D power spectrum, so-called power spectrum tomography, would rely on
the availability of photometric redshift estimates for the galaxies
involved in the analysis (e.g. Hu 1999; 2002).  Croft \& Metzler
(2000) demonstrated that the intrinsic alignment signal is more severe
when correlations within a narrow slice in source galaxy redshift
space are considered. The application of a redshift dependent filter,
similar to the one we have described here, would greatly suppress the
intrinsic alignment systematic.  Instead of slicing galaxies in
redshift, and calculating the correlation function for galaxies in the
same redshift bin -- which both increases the relative importance of
intrinsic alignments and reduces the total number of pairs -- one
should correlate galaxies from different redshift bins $i,j$, where the
bin width would be chosen to be of the same order as the redshift
uncertainty. This would yield a set $\xi_{i,j}(\theta)$ of
correlation functions for which the intrinsic signal is strongly
suppressed for $i\ne j$. When comparing these correlation functions
with predictions from cosmological models, their covariance must be
taken into account, which may turn out to be fairly complicated (see
Schneider et al.\ 2002b for the case with no redshift information).
This redshift slicing is most straightforwardly done with the
correlation function, although for other estimators of the two-point
cosmic shear statistics, such as the shear dispersion or the aperture
mass, similar pair redshift-dependent estimators are easily
constructed.

Most of what has been said above also applies to higher-order cosmic
shear statistics. It is well known that the three-point statistics,
i.e. the skewness, contains very useful cosmological information
(e.g. van Waerbeke et al.\ 1999). As true for the 2-point statistics,
the three-point correlation function is the quantity which is most
easily derived from a cosmic shear survey. Bernardeau et al.\ (2002a)
have constructed a statistics based on the measured three-point
correlation function (see Schneider \& Lombardi 2002 for the
classification of three-point shear correlators), and Bernardeau et
al.\ (2002b) obtained a significant detection in the VIRMOS-DESCART
survey data. It is unclear whether, and by how much, intrinsic
ellipticity correlations affect these measurements. Furthermore, up to
now no B-mode estimator in the three-point function has been devised
which may indicate the presence of intrinsic alignment
effects. Redshift slicing can of course also be done for the
three-point function which may be the only way to measure these lensing
statistics without the potential influence of intrinsic effects; one
needs to suppress those triplets of galaxies where the probability of
all three being at the same redshift is not negligibly small (i.e.,
the estimator is unaffected by intrinsic effects if two of the three
are at the same redshift).

Of course, the intrinsic alignment signal that we seek to suppress
when performing a lensing analysis is also interesting in its own
right, since it places constraints on the formation and evolution of
galaxies. Suitable catalogues of galaxies to quantify this signal, and
its evolution with redshift, will be a natural by-product of large
cosmic shear surveys with photometric redshift information. For
example, using a filter for the determination of the ellipticity
correlation as in (\ref{est}) with ${\cal Z}'_{i,j}=1-{\cal Z}_{i,j}$
will make this estimate dominated by the intrinsic alignments, and
could thus be used as a first estimate of it. More sophisticated
techniques would include the consideration of the resulting
correlation function in dependence of the width of the weighting
function, and extracting the lensing and intrinsic signal from this
functional form.

If the intrinsic alignment of galaxies indeed occurs, and in
particular if the intrinsic effect is as strong as suggested by some
models (e.g. Jing 2002), a deep multi-colour wide-field survey with a
broad range of filters will be necessary, and rewarding, to remove
this systematic from cosmic shear measurements. Most likely, the
near-IR imaging will present the bottleneck, limiting the magnitude -
and thus the effective number density - of galaxies that can be used
for cosmic shear. A wide-field near-IR camera in space, such as the
PRIME satellite mission, would be an ideal supplement to the planned
extensive ground-based optical cosmic shear surveys.

After we had completed this paper, we became aware of work by Heymans
\& Heavens (2002), also addressing how redshift information can be
used to reduce the contamination from intrinsic alignments. They apply
their technique to estimate the reduction of the intrinsic signal in 
several surveys, including the SDSS photometric and spectroscopic samples.

\begin{acknowledgements}
We would like to thank Doug Clowe, Thomas Erben, Martin Kilbinger,
Patrick Simon, and in particular Marco Lombardi and Ludo van Waerbeke
for useful discussions. We would also like to thank the anonymous
referee 
for very helpful comments. This work was supported by the Deutsche
Forschungsgemeinschaft under the project SCHN 342/3--1 and by the TMR
Network ``Gravitational Lensing: New Constraints on Cosmology and the
Distribution of Dark Matter'' of the EC under contract
No. ERBFMRX-CT97-0172.
\end{acknowledgements}

\end{document}